# Robustness under functional constraint: The genetic network for temporal expression in *Drosophila* neurogenesis

Akihiko Nakajima[1], Takako Isshiki[2], Kunihiko Kaneko[1,3], Shuji Ishihara[1,4]

1. Department of Basic Science, University of Tokyo, 3-8-1 Komaba, Meguro-ku, Tokyo 153-8902, Japan

2. Center for Frontier Research, National Institute of Genetics, 1111 Yata, Mishima, Shizuoka 411-8540, Japan

3. ERATO Complex Systems Biology Project, Japan Science and Technology Agency, 3-8-1 Komaba, Meguro-ku, Tokyo 153-8902, Japan

4. PRESTO, Japan Science and Technology Agency, 4-1-8 Honcho, Kawaguchi, Saitama 332-0012, Japan

To whom the correspondence: Akihiko Nakajima

Email address: nakajima@complex.c.u-tokyo.ac.jp

Telephone/fax: +81 3 5454 6732.

**Abstract**

Precise temporal coordination of gene expression is crucial for many developmental processes. One central question in developmental biology is how such coordinated expression patterns are robustly controlled. During embryonic development of the *Drosophila* central nervous system, neural stem cells called neuroblasts sequentially express a group of genes in a definite order, which generates the diversity of cell types. By producing all possible regulatory networks of these genes and examining their expression dynamics numerically, we identify requisite regulations and predict an unknown factor to reproduce known expression profiles caused by loss-of-function or overexpression of the genes *in vivo*, as well as in the wild type. We then evaluate the stability of the actual *Drosophila* network for sequential expression. This network shows the highest robustness against parameter variations and gene expression fluctuations among the possible networks that reproduce the expression profiles. We propose a regulatory module composed of three kinds of regulations which is responsible for precise sequential expression. The present study suggests an underlying principle on how biological systems are robustly designed under functional constraint.

**Introduction**

Precise coordination of cell fate decisions is crucial in the development of multicellular organisms. In the developmental processes, where a series of events occurs at a specific place and time, gene regulatory networks are responsible for implementing the reliable biological functions [1,2]. In order to obtain the system-level understanding of the processes, it is necessary to integrate the molecular machinery of each regulation with the architecture and dynamics at the regulatory network level. Biological functions achieved by gene networks are generally expected to possess robustness, i.e., insensitivity of system properties against a variety of perturbations that may be originated from fluctuations during development and mutations through evolution. Recent investigations have addressed the questions on how robust functions in gene or signaling networks are achieved through underlying network architecture and its dynamical properties [3,4,5,6,7]. An illustrative example in developmental systems on this subject is segmentation of *Drosophila melanogaster*, which has been studied both experimentally and theoretically [8,9,10]. The requisite regulations or architecture of this system have been discussed from the point of network level description [10,11,12,13,14], and it is suggested that the underlying gene network is designed to perform the process in a robust manner [15,16,17].

Besides spatial patterning, temporal patterning also plays important roles in various developmental processes [18,19,20]. One of the most studied systems is the development of the *Drosophila* central nervous system (CNS), in which the sequential expression of genes coordinates cell-fate decisions. The neural stem cells called neuroblasts (NBs) express a series of transcription factors in a definite order: Hunchback (Hb), Krüppel (Kr), Pdm1/Pdm2 (Pdm), and Castor (Cas) (Fig. 1A to C) [21,22,23,24]. In addition, the fifth factor Seven-up (Svp) is expressed in the time window between Hb and Kr expression [25]. In association with this sequential expression, NBs divide asymmetrically to bud off a series of ganglion mother cells (GMCs). Each GMC undergoes an additional division to generate typically two postmitotic neurons. The sequentially expressed transcription factors in NBs control temporal specification of cell fate of neurons, thereby establishing the diversity of cell types in the *Drosophila* CNS.

Isolated NBs exhibit sequential expression *in vitro* and differentiate into various neurons in the same manner as *in vivo* [26,27]. Hb expression is switched off by Svp in a mitosis-dependent manner, while the subsequent expression of Kr, Pdm and Cas proceeds in a mitosis-independent manner [25,28]. These observations suggest that sequential expression of the genes is regulated cell-autonomously and occurs through mutual interactions among the factors.

In this study, we address the robustness of the gene network for sequential expression in the *Drosophila* CNS. One of the promising approaches to characterize robustness of biological systems is to compare the actual network with other possible network architectures with respect to function and robustness. Wagner considered how network architecture and robustness are related by studying circadian oscillation networks [29], although these networks lack the direct biological counterpart. Ma *et al*. studied the architecture of the *Drosophila* segmentation network [30], in which they had to arbitrarily eliminate components to reduce the size of the entire network. From theoretical and computational points of view, one advantage of studying temporal patterning in the *Drosophila* CNS is that the number of system components is so small that we can perform a comprehensive analysis of network architecture without any loss of biological relevance.

First, we explored the conditions necessary for gene regulatory networks to reproduce the observed expression patterns in both wild type (WT) and mutants. We did not confine ourselves to only known regulations for sequential expression, but rather searched all possible networks that could reproduce the observed expression patterns. Studying the common structure of the specified genetic networks, we detect requisite regulations and predict an unknown factor to reproduce known expression profiles. Second, we compared these functional networks with the actual *Drosophila* network in terms of network architecture and robustness of the expression pattern. We found that the *Drosophila* network is highly robust and stable among possible functional networks. We discuss how the architecture of the *Drosophila* network implements robustness of sequential expression against both cell-to-cell variations and intracellular fluctuations.

**Results**

**Temporal patterning network of *D. melanogaster* NBs**

Experimentally reported expression profiles of the temporal transcription factors are summarized in Figure 1D for WT, loss-of-function, and overexpression embryos [22,23,25,27,31,32]. These sequential expressions are considered to be produced (or at least modulated) by mutual regulations among the temporal transcription factors [21,22]. We reconstructed the gene network for sequential expression in *Drosophila* NBs from the literature as shown in Figure 1E and F (for references, see Table I).

**Modeling gene network dynamics by Boolean description**

First, we considered the necessary conditions for the network architecture to reproduce the sequential expression patterns of both WT and mutants. To investigate gene expression dynamics, we adopted a Boolean-type model [6] (see **Materials and methods** for details of the model and the following analysis),

$$X_i^{t+1} = f_i(\{X_j^t\}) = \begin{cases} 1 & (\sum_j J_{ij} X_j^t > 0) \\ 0 & (\sum_j J_{ij} X_j^t < 0) \\ \theta_i & (\sum_j J_{ij} X_j^t = 0) \end{cases} \quad (1),$$

where $X_i^t$ represents the expression state of gene $i$ ($i \in$ {*hb*, *Kr*, *pdm*, *cas*}) at the *t*-th time step and takes either 1 (ON) or 0 (OFF). Regulation from gene *j* to gene *i* is either positive ($J_{ij} > 0$), negative ($J_{ij} < 0$), or zero ($J_{ij} = 0$), which corresponds to activation, repression, or absence of regulation, respectively. The state of gene *i* at the next step ($X_i^{t+1}$) is 1 when the sum of the regulatory inputs is positive ($\sum_j J_{ij} X_j^t > 0$) or 0 when the sum is negative ($\sum_j J_{ij} X_j^t < 0$). When the sum equals zero ($\sum_j J_{ij} X_j^t = 0$), $X_i^{t+1}$ takes the default expression state of the gene $\theta_i$: $\theta_i \in \{0, 1\}$. In this study, the value of $J_{ij}$ is supposed to take one of the discrete values $J_{ij} \in \{1, 0, -5\}$. The large negative value of $J_{ij}$ signifies that the expression of a gene is completely shut off in the presence of a repressor. Initial states of the genes are set to 0 except for Hb, which emulates the expression state of NBs in the first stage of the sequential expression [21,22]. Thus far, the only known function of Svp during the early stage is downregulation of Hb. There is no evidence that Svp regulates or is regulated by other temporal transcription factors during the expression series: Kr → Pdm → Cas [25]. In addition, Hb is only regulated by Svp and not by the other three factors (Kr, Pdm, and Cas). Thus, in the model, we assumed a pulsed expression of Svp as an input to the system, resulting in downregulation of Hb at the next time step. The temporal expression dynamics of Kr, Pdm, and Cas follow Eq. (1) with assigned values of $J_{ij}$ (Fig. 1F).

**The regulatory networks of known factors do not reproduce the experiments**

Based on the above formulation, we investigated whether the reconstructed *Drosophila* gene network (Fig. 1E and F) is enough to reproduce the sequential expression observed in WT, as well as all of the known single loss-of-function and overexpression mutants: *hb*$^-$, *Kr*$^-$, *pdm*$^-$, *cas*$^-$, *hb*$^{++}$, *Kr*$^{++}$, *pdm*$^{++}$, and *cas*$^{++}$ (here, "++" means overexpression of the gene) (Fig. 1D, Table II). At the moment, we cannot specify the values of the parameters $\theta_{Kr}$, $\theta_{pdm}$, and $\theta_{cas}$ from empirical data, thus each value could be arbitrarily chosen from $\theta_i \in \{0, 1\}$ ($i \in \{Kr, pdm, cas\}$). We studied all $2^3$ combinations of $\{\theta_i\}$ and found that the dynamics coincide with the expression profile in WT, but do not in some of the mutants for each choice of the parameters (examples shown in Fig. 2). Depending on the parameter values, the expression dynamics changed to some extent, but none of the possible combinations reproduced the expression profiles of all of the mutants. For example, in the case of $\theta_{Kr} = 0$, $\theta_{pdm} = 0$, and $\theta_{cas} = 1$, the dynamics of the network for $hb^-$ and $Kr^-$ did not agree with the experiments (Fig. 2A), while in the case of $\theta_{Kr} = 1$, $\theta_{pdm} = 1$, and $\theta_{cas} = 1$, the dynamics of $hb^-$ and $pdm^-$ did not (Fig. 2C).

We then investigated whether other networks than the *Drosophila* network can reproduce the observed expression profiles by checking all the $3^{12}$ (=531,441) combinations of $J_{ij}$ values. The dynamics agreed with the expression profile in WT for a large number of networks (39,391 out of 531,441), but any networks composed of *hb*, *Kr*, *pdm*, *cas*, and *svp* did not reproduce the profiles in both WT and mutants.

**Introduction of a presumptive factor is sufficient to reproduce the expression profiles**

Preceding results indicate the difficulty of reproducing the observed expression patterns only with known constituents. We therefore introduced an additional presumptive regulator (*x*). The expression state of *x* was assumed to start in the ON state and change into OFF, or *vice versa* at $t = \tau_{switch}$ ($0 \leq \tau_{switch} \leq \tau_{end}$) (see **Materials and methods**). Including this assumption we reinvestigated the dynamics of all $3^{15}$ (=14,348,907) possible regulatory networks with all the possible switching timing of *x*. In the case that the expression of *x* switches OFF to ON, none of the networks conformed to the expected expression profiles. On the other hand, in the case that the expression of *x* switches ON to OFF, we found that 384 networks (<0.003%) reproduced the expression profiles of both WT and mutants. We refer to the detected networks as "the functional networks" in the rest of this study.

We compared the network architectures and found that the regulations shared among all the functional networks are coincident with experimentally verified regulations (colored as black in Fig. 3A). In addition, activation of *Kr* and repression of *cas* by a presumptive factor *x* appear in all of the functional networks (colored as brown in Fig. 3A). Therefore, we conclude that the genetic network composed of these common regulations is a minimum network that is necessary and sufficient to reproduce the expression profiles of WT and mutants.

To quantify the similarity among the functional networks, we measured the distances of the 384 functional networks from the actual *Drosophila* network (Fig. 3C); The distances are defined by the number of different regulations (see **Materials and methods**). As a reference, we also performed the same analyses of distance measurement for all possible networks and the networks that are randomly reconnected from functional networks (see **Materials and methods**). For all possible networks, the frequency distribution of the distances shows that the network architectures are different from the actual *Drosophila* network by 7.8 ± 1.5 regulations. The reconnected networks yield similar results albeit with slightly decreased distances (7.0 ± 1.7 regulations). In contrast, the architectures of the functional networks differ by only 2.4 ± 1.1 regulations. The architectures of the functional networks resemble that of the actual *Drosophila* network. These indicate the gene networks that reproduce the known sequential expression patterns are highly constrained in their topologies.

**Robustness of the *Drosophila* network against parameter variations and expression noise**

Since there are multiple network architectures that explain the observed expression profiles as shown above, we then ask the characteristics of the actual *Drosophila* network among the functional networks. From the biological point of view, the sequential expression in NBs should proceed reliably despite developmental disturbances such as cell-to-cell variation and intracellular fluctuations. We thus evaluated the stability of sequential expression for each of the detected functional networks and compared the properties of the actual *Drosophila* network to those of the other networks. To address the problem quantitatively, we extended the previous Boolean-model into a model of ordinary differential equations with fluctuations in gene expression, where the concentrations of mRNAs $\{M_i(t)\}$ and proteins $\{P_i(t)\}$ obey the following equations [33,34] (see **Materials and methods** for the details of the model and the following analysis):

$$\frac{dM_i(t)}{dt} = \gamma_M \left[ F_i(\{P_j(t)\}) - M_i(t) \right] + \xi_i(t),$$
$$\frac{dP_i(t)}{dt} = \gamma_P \left[ M_i(t) - P_i(t) \right]. \tag{2}$$

Here $i$ refers to one of each gene: $i \in \{hb, Kr, pdm, cas, x\}$. The variables $\{M_i(t)\}$ and $\{P_i(t)\}$ take continuous values, unlike the previous Boolean description. The precise function form of promoter activities $\{F_i(\{P_j(t)\})\}$ is dependent on the regulatory interactions of the genetic networks $\{\tilde{J}_{ij}\}$ and the default promoter activities $\{S_i\}$, corresponding to the Boolean model. The time-dependent variables $\{\xi_i(t)\}$ represent the noise in promoter activities with intensity $\sigma$.

Typical dynamics of the *Drosophila* network are shown in Figure 4, where sequential expression of WT is reproduced. The dynamics of the model are largely dependent on the parameter values and the noise intensities, and coincide with the experimental observations only under appropriate conditions. Therefore, such sensitivity to parameter variation is important for the development to proceed under environmental and individual fluctuations.

To characterize sensitivity, we measured the fraction of successes, that is the fraction of the parameter sets that can reproduce the expression profile of WT among all the trials of random parameter assignments [15,30]. To obtain the effect of parameter variation, we carried out the simulation without stochastic terms in Eq. (2) (i.e., $\sigma = 0$). In each network, we repeated the simulations with random assignment of parameter values and calculated the fraction of successes (Fig. 5A). The *Drosophila* network scored the highest fraction of successes among the functional networks, and the networks closer to the *Drosophila* network tended to have higher scores.

We also investigated the dynamical stability of the gene networks against fluctuations. In this case, we performed the stochastic simulations in Eq. (2) with finite $\sigma$. To evaluate stability against noise, we chose the parameter values with which the expression profile is reproduced in the absence of noise. We then measured the relative fraction of successes under fluctuation. As is shown in Figure 5B, the fraction of successes under expression noise increased with the similarity to the actual *Drosophila* network as the fraction of successes under parameter variations. Thus, the *Drosophila* network lies at the top level of the functional networks in terms of robustness against these perturbations.

**Regulations that heighten functional stability**

Because the *Drosophila* network has several additional regulations further to the minimum functional network (gray arrows in Fig. 3A), these regulations may be responsible for the robustness shown above. We compared the robustness among the networks with or without the additional regulations. The fraction of successes against parameter variations for these networks are plotted in Figure 6A. The minimum network reproduces the sequential expression under the appropriate parameters, but the robustness is much lower than that of the *Drosophila* network. The scores of networks that lack one of the regulations fall between the minimum and the *Drosophila* network.

Stability to expression noise was also evaluated by changing noise intensity, and similar results were obtained (Fig. 6B). The fraction of successes decreases as the noise intensities get larger, but the effect of noise on the *Drosophila* network is less severe than that on the minimum network. Thus each of these regulations contributes to robustness of the system.

To elucidate the roles of these regulations, we tried random parameter assignments for each of these networks and sampled successful parameter sets that reproduce WT sequential expression profile (Fig. 7). In the *Drosophila* network (Fig. 7A), wide ranges of parameter values are allowed, indicating that this network reproduces the required profile without quantitative tuning of parameters and thus shows high robustness. For the other networks (Fig. 7B to E), the ranges are narrower for some parameters (as clearly seen in $S_{pdm}$ and $S_{cas}$), and the numbers of successful parameter sets are less than those for the *Drosophila* network.

How is the robust nature of the *Drosophila* network implemented by these regulations? As seen above, the parameter values of $S_{pdm}$ and $S_{cas}$ (default promoter activities of *pdm* and *cas*) are most influenced by the loss of these regulations. Because expression of a gene is induced by either the default promoter activity or the activators (see **Materials and methods**), additional regulations in the *Drosophila* network (gray arrows in Fig. 3A) may compensate for the loss of default activities. To verify this possibility, we measured the dependence of the fraction of successes on the strength of regulations ($\tilde{J}_{pdm, Kr}$, $\tilde{J}_{cas, pdm}$, and $\tilde{J}_{cas, hb}$) and of default promoter activities ($S_{pdm}$ and $S_{cas}$) (Fig. 8A to C).

Fig. 8A shows the fraction of successes for random assignments of parameter values under given strengths of $\tilde{J}_{pdm, Kr}$ and $S_{pdm}$. To score high reproducibility, $S_{pdm}$ must be large for small $\tilde{J}_{pdm, Kr}$, but need not for sufficiently large $\tilde{J}_{pdm, Kr}$. This indicates that activation of *pdm* expression by *Kr* indeed compensates for the loss of default promotor activity of *pdm*. Thus, for the network lacking this regulation, the default promoter activity is necessary because inductions from other factors are absent. A similar relationship is found between $\tilde{J}_{cas, pdm}$ and $S_{cas}$ (Fig. 8B).

As for repression of *cas* by *hb*, the role for robustness seems to be different from the above two. When the absolute value of $\tilde{J}_{cas, hb}$ is small, $S_{cas}$ must be small to achieve a high fraction of successes (Fig. 8C). As $|\tilde{J}_{cas, hb}|$ becomes larger, a higher value of $S_{cas}$ is allowed. This is because the repression from *hb* to *cas* reduces the expression of *cas* in the early stage of sequential expression. Thus the existence of this regulation contributes to the robustness against the parameter variation of $S_{cas}$. Grosskortenhaus *et al*. suggested the direct repression from *hb* to *cas* [23], although there is no confirmative evidence to our knowledge. If any, this regulation would contribute to the robustness of the system.

**Discussion**

Through the present analyses, we obtained 384 functional networks that reproduce the sequential expression of both WT and mutants. The detected functional networks exhibit high similarity in regulatory interactions among the transcription factors (Fig. 3). This exemplifies the importance of the regulations in the minimum network for the sequential expression. In addition, the actual *Drosophila* network scores quite high on reproducibility of the WT sequential expression among all the functional networks (Fig. 5 and 6). Below, we discuss the biological implications of the temporal patterning of *Drosophila* NBs drawn from our numerical analyses.

**Two regulatory interactions from a presumptive factor are necessary and sufficient to reproduce the expression patterns of WT and mutants**

In this study, we introduced an additional presumptive factor *x* to obtain the networks that reproduce the sequential expression of both WT and mutants. As *x* is hypothetical, we discuss its validity here.

Since the loss-of-function mutant of any one gene has only minor effects on the expression sequence (Fig. 1D), several previous reports suggested the existence of either unknown regulators or an additional clock mechanism for driving the sequential expression [22,23]. Our assumption is feasible one for explaining experimental results in that it does not need any other clock mechanism or superfluous multiple regulators. It is notable that our analysis indicates that the possible regulations of the presumptive factor are highly restricted; the expression of *x* switches ON state to OFF (Fig. 4), and all the functional networks have activation of *Kr* and repression of *cas* by *x* (Fig. 3A). Thus, our assumption is testable in future experiments *in vivo*.

We should note that although regulator *x* is necessary to explain the mutant profiles, the mutual regulations of known factors are sufficient to reproduce the WT sequential expression (Fig. 1D). Therefore, the regulations among *hb*, *Kr*, *pdm* and *cas* would play a primary role as discussed below.

**Minimum network for the sequential expression**

An effective way to capture network function is to focus on the specific substructures (network motifs or modules) [1,13,14,16,30,35]. Comparing all the functional networks, we detected the minimum structure for the sequential expression, which contains two successive regulatory loops (Fig. 3A and 9A); one is composed of *hb*, *Kr*, and *pdm*, and the other of *Kr*, *pdm*, and *cas*. In each loop, one gene represses previous and second next factor. The repressions of the second next factors (*hb* to *pdm* and *Kr* to *cas*) define the induction timing of the regulated factors, since they are kept repressed until the regulators are switched off. The feedback repression of the previous factors (*pdm* to *Kr* and *cas* to *pdm*) ensures their downregulation, which promotes the progress of the sequential expression. These coincide with the observation by Kambadur *et al.*, who showed experimentally that the repressions from *hb* and *cas* define the temporal window of Pdm [21]. These repressive regulations and the activation from *hb* to *Kr* compose the minimum network for sequential expression (Fig. 9A). Although they are enough to reproduce the sequential expression under appropriate conditions, the expression profiles could be easily perturbed by the parameter variations or the increase of noise (Fig. 5 and 9A).

**Robustness of the *Drosophila* network: mechanism generating the precise sequential expression**

In the two loops of the *Drosophila* network, the activations from one gene to the next (*Kr* to *pdm* and *pdm* to *cas*) exist in addition to the repressive regulations. Other functional networks do not necessarily have these activations, but the activations can compensate for the loss of default promoter activities (Fig. 8A and B). These regulations achieve precise expression by enhancing the correlations among the factors and heightening the stability against fluctuations (Fig. 5B and 6B). From these results, we conclude that three different kinds of regulations (the activation of next factor, feedback repression, and repression of second next one) compose a regulatory module for precise temporal expression as summarized in Figure 9B. The feature of this network module embodies the robustness of the *Drosophila* network.

Do above discussions have any implications on other developmental processes? In the studies of spatial patterning in *Drosophila* segmentation, it was claimed that the frequent substructure feed forward loop (FFL) can set the positions of expression domains [13], and mutual feedback repressions between the gap genes also have a pivotal role for the formation of expression domains with steep boundaries [12,35]. In the case of the *Drosophila* network for sequential expression, preceding genes activate the next ones, while these genes repress the preceding ones. Similar regulatory interactions are reported in the yeast cell cycle by Lau et al. [36]. Thus, such asymmetric mutual regulations would be a general mechanism that serves as precise switches in the process of temporal patterning.

**Role of the robustness in *Drosophila* neurogenesis**

We showed that the *Drosophila* network contains not only the regulations necessary for generating sequential expression, but also additional ones to achieve higher precision in the expression. In each hemisegment of *Drosophila* embryo, 30 different NBs are generated through spatial heterogeneity [37]. To guarantee *Drosophila* NBs sequentially express common temporal transcription factors despite their differences, the robustness of the system may become important.

The robust nature of the *Drosophila* network could be the consequence of evolutionary optimization in the reproducibility of the sequential expression under functional constraint. In future, we expect that experimental manipulation of corresponding enhancers will be able to clarify the relevance of each regulation to the temporal patterning and stability.

**Materials and methods**

**Analysis of temporal dynamics of the genetic networks with the Boolean model**

Here we describe the details of the Boolean model (Eq. (1)). The expressions of *svp* and *x* occur as inputs to the system. A pulse of *svp* expression always occur at $t = 1$. Expression of *x* switches either from ON to OFF state, or from OFF to ON at $t = \tau_{switch}$ ($0 \leq \tau_{switch} \leq \tau_{end}$). Once we assigned the switching time of *x* expression ($\tau_{switch}$), its value had been fixed through the analysis of expression patterns for all the genotypes. Because the autonomous pulsed expression of *svp* results in *hb* downregulation, we set $J_{hb, svp} = -5$, $J_{hb, j} = 0$ ($j = hb$, *Kr*, *pdm*, *cas*, or *x*), and $J_{k, svp} = 0$ ($k = Kr$, *pdm*, or *cas*) throughout this study. The time step at which we finish the simulation ($\tau_{end}$) was set as $\tau_{end} = 12$.

We thus investigated the behaviors of the remaining three factors (*Kr*, *pdm*, and *cas*) under the given regulatory interactions $\{J_{ij}\}$. The total number of combinations of the parameters is $3^M \times 2^3$ (the number of possible network architecture $\{J_{ij}\}$ multiplied by the number of default expression states $\{\theta_i\}$), where *M* is the number of regulations. To simulate the dynamics for mutants, we always set the expression state of the corresponding gene to 0 (OFF) for loss-of-function or 1 (ON) for overexpression. We then examined whether the temporal dynamics of the genetic networks are coincident with the expression profiles of each mutant (Fig. 1D and Table III).

**Analysis of network statistics**

In order to measure the similarity between the functional networks and the actual *Drosophila* network, we used two types of network ensembles as references. One is the ensemble of the possible network architectures. The other is a set of reconnected networks generated from the functional networks by iterative random reconnections of the matrix elements (1000 iterations). The numbers of positive and negative regulations are preserved in the iterations.

To count the number of different regulations between functional networks and the actual *Drosophila* network, we neglected the regulations from *x*, and positive self-feedbacks because the existence of those is uncertain from the experimental data.

**Continuous model of the expression dynamics**

We introduced the continuous model with stochasticity as shown in Equation (2). The promoter activity of gene *i* (*i* = *hb*, *Kr*, *pdm*, *cas*, or *x*) is described as follows,

$$F_i(\{P_j(t)\}) = \frac{\left[g(S_i + \sum_j \tilde{J}_{ij} P_j)\right]^\alpha}{K_M^\alpha + \left[g(S_i + \sum_j \tilde{J}_{ij} P_j)\right]^\alpha}.$$

Regulatory interactions $\{\tilde{J}_{ij}\}$ are continuous equivalents of $\{J_{ij}\}$ in the Boolean model, and $g(x)$ is a piece-wise linear function such that $g(x) = x$ for $x > 0$, and $g(x) = 0$ for $x < 0$. The parameters $\{S_i\}$ are the default activities of the promoters. Transcription of a gene is induced when the total regulatory inputs become positive ($S_i + \sum_j \tilde{J}_{ij} P_j > 0$), and is suppressed when they become negative ($S_i + \sum_j \tilde{J}_{ij} P_j < 0$). In order to consider the effect of fluctuations on the expression dynamics, we introduced additive white

Gaussian noise $\{\xi_i(t)\}$: $\langle \xi_i(t)\xi_j(t')\rangle = \sigma_i^2 \delta_{ij}\delta(t-t')$ (Eq (2)), where $\sigma_i$ is the noise intensity of gene $i$.

The expressions of *hb* and *x* are induced only by the default promoter activities because all the regulations are absent for these two ($\{\tilde{J}_{hb,i}\}=\{\tilde{J}_{x,i}\}=0$). To describe the expression change of *hb* and *x*, the promoter activities of these two are set as $S_{hb} > 0$ for $t < \tau_{hb,\text{off}}$ ($S_x > 0$ for $t < \tau_{x,\text{off}}$), and $S_{hb} = 0$ for $t > \tau_{hb,\text{off}}$ ($S_x = 0$ for $t > \tau_{x,\text{off}}$). The promoter activities of the others are always assumed to exist ($S_{Kr}$, $S_{pdm}$, and $S_{cas} > 0$). The noise intensities are also set as $\sigma_i = \sigma$ (>0) for $t < \tau_{i,\text{off}}$ and $\sigma_i = 0$ for $t > \tau_{i,\text{off}}$ ($i = hb, x$). Those of the other genes are $\sigma_j = \sigma$ (>0) ($j = Kr, pdm, cas$), Here we simply assume that the noise intensities of the genes take the same value $\sigma$.

**Analysis of the robustness of the networks**

For the continuous model, we considered two different types of robustness: the reproducibility of the sequential expression against parameter variations and dynamical stability against temporal fluctuations. To analyze the former, the default promoter activities $\{S_i\}$ were assigned randomly within the defined ranges. The values of the matrix $\{\tilde{J}_{ij}\}$ were set to 0 when the corresponding regulations were absent (The corresponding element of the Boolean model takes as $J_{ij} = 0$) or assigned randomly when they are present ($J_{ij} \neq 0$). In order to confine our attention to the properties of network architectures, the other parameters ($\gamma_M$, $\gamma_P$, $K_M$, and $\alpha$) were fixed throughout the analysis. The ranges and the fixed values of the parameters are listed in Table IV.

Robustness against temporal fluctuations is measured as explained in the main text.

To judge whether the dynamics coincide with the expression profile in *Drosophila* NBs, the dynamics of the protein concentrations $\{P_i\}$ were discretized to 1 (0) for $P_i > P_{th}$ ($P_i < P_{th}$). The threshold $P_{th}$ was set as $P_{th} = 0.2$. The temporal dynamics of a network were accepted when the discretized dynamics satisfied the condition for WT in Table III.

In the simulations, we found that the existence of positive self-regulation enhanced the fraction of successes in many cases, but hardly affected the sequential expression. To focus on the contributions of mutual regulations of genes to robustness, we neglected the positive self-feedback regulations and confined the analysis to 120 out of 384 functional networks.

**Acknowledgements**

We thank C. Q. Doe for his kind permission to reprint the published data for Figure 1B in this article. We also thank H. Takagi and T. Shibata for discussion, and T. Suzuki and K. Fujimoto for useful comments on the manuscript. This study was supported by a Grant-in-Aid for JSPS Fellows (to A.N.) and by PRESTO of JST (to S.I.).

**Figure legends**

**Figure 1. Sequential expression of temporal transcription factors within neuroblasts in the *Drosophila* CNS.**

**(A)** The relative position of neuroblasts (NBs) in *Drosophila* embryo. The picture is the ventral view of NBs and shows Cas expression in the NBs at developmental stage 12. The bracket indicates a single segment. Dashed line corresponds midline. Scale bar: 40μm. **(B)** The expression levels of Hb, Kr, Pdm, and Cas in a single NB (NB 2-4 lineage) are shown from the developmental stage 10 to 12: early stage 10 (st. 10); early stage 11 (e11); mid stage 11 (11); late stage 11 (l11); mid stage 12 (12); late stage 12 (l12). The pictures are partially reprinted from [22]. **(C)** Schematic representation of the change of the expression pattern in a single NB. **(D)** The expression profiles of WT, loss-of-function and overexpression mutants of the genes observed in the experiments (for references, see Table II). **(E)** Reconstructed genetic network for sequential expression in *Drosophila* NBs. Repression from *hb* to *cas* (dashed line) was suggested to exist [23], although there is no direct verification. When the *Drosophila* network is invoked in this article, this regulation is also included. **(F)** Matrix representation of the *Drosophila* network.

**Figure 2. The reconstructed *Drosophila* network cannot reproduce the experimentally reported expression profiles.**

Sequential gene expression of reconstructed *Drosophila* network is simulated using Boolean model. The grids filled with colors represent ON states of the genes. The dynamics could be different depending on the choice of the default expression states $\{\theta_i\}$. **(A)** $\theta_{Kr} = 0$, $\theta_{pdm} = 0$, and $\theta_{cas} = 1$, **(B)** $\theta_{Kr} = 0$, $\theta_{pdm} = 1$, and $\theta_{cas} = 1$, and **(C)** $\theta_{Kr} = 1$, $\theta_{pdm} = 1$, and $\theta_{cas} = 1$.

**Figure 3. Architecture of the detected functional networks.**

**(A)** Architecture of the functional networks reproducing the gene expression profiles observed in the experiments. The black arrows are the regulations that appear in all the functional networks. The brown arrows are the regulations from the presumptive factor *x* that also appear in all the functional networks. The other regulations existing in the actual *Drosophila* network are shown by gray arrows. **(B)** Matrix representation of the functional networks. Elements of $\{J_{ij}\}$ are shown as either + for activation, – for repression, or 0 for the absence of regulation. **(C)** Frequency distributions of the distances of networks from the *Drosophila* network. The distributions are drawn from the functional networks ($N = 384$; magenta), all the possible networks ($N = 14{,}348{,}907$; blue), and the networks randomly reconnected from the functional ones ($N = 38{,}400$; yellow). From each of the functional networks, 100 reconnected networks were generated. The regulatory interactions from *x* and positive self-feedbacks are neglected in counting the number of different regulations.

**Figure 4. Temporal dynamics of the *Drosophila* network in the continuous model.**
The dynamics of expression levels of proteins $\{P_i(t)\}$ with different parameter values (upper) and discretized representation of a typical temporal dynamics (lower). In addition to the known genes, the presumptive factor $x$ is also incorporated. The expression level of X changes from high level to low as in the previous model. Each gene is considered to be in the ON state when the expression level is larger than a threshold $P_{th}$. The parameter values of $\{\tilde{J}_{ij}\}$ and $\{S_i\}$ are randomly selected from the following ranges: $|\tilde{J}_{ij}| \in [10^{-1}, 10^0]$ for $\tilde{J}_{ij} > 0$, $|\tilde{J}_{ij}| \in [10^0, 10^1]$ for $\tilde{J}_{ij} < 0$, $S_{hb}$, $S_{Kr}$, $S_x \in [10^{-2}, 10^0]$, and $S_{pdm}$, $S_{cas} \in [10^{-1}, 10^0]$. Noise intensity is set as $\sigma = 0.05$. The other parameter values are set as shown in Table IV.

**Figure 5. The robustness of the gene expression profiles in the functional networks.**
**(A)** The fraction of trials that reproduce the experimental expression profile against random assignments of parameters. The values of $\{\tilde{J}_{ij}\}$, $\{S_i\}$, and $\tau_{x,off}$ are randomly chosen within the ranges shown in Table IV. The other fixed parameter values are also listed in Table IV. Neglecting the positive self-feedback regulations in the 384 functional networks, 120 networks were chosen and investigated (**Materials and methods**). The dynamics were checked for 50,000 trials in each network. The networks were sorted based on the distance from the *Drosophila* network. Because there are a few possible regulations from the unknown factor $x$, the networks with $N_d = 0$ exist more than one. **(B)** The fractions of the trials that reproduce the experimental profile under expression noise (vertical axis) are plotted against the fraction of successes against the random parameter assignments. To analyze the stability against noise, we used 1000 different parameter sets with which the expression profile is reproduced in the absence of noise for each networks. The dynamics were checked for 50 trials for each parameter set. Noise intensity is set as $\sigma = 0.08$.

**Figure 6. Contribution of the actual regulations to the robustness of the system.**

**(A)** The fraction of the trials that reproduce the experimental WT expression against parameter variations. The data of Figure 5A are replotted for the *Drosophila* network, the networks lacking an indicated regulation (one of the gray arrows in Fig. 3A), and the minimum network (black and brown arrows in Fig. 3A). **(B)** The fractions of the trials that reproduce the experimental profile under the gene expression noise with various intensities. We used 5,000 different parameter sets with which the profile is reproduced in the absence of noise. The dynamics are checked for 50 trials for each parameter set.

**Figure 7. Graphical representation of parameter sets with which the WT sequential expression profile is reproduced.**

**(A)** the *Drosophila* network, the networks lacking **(B)** activation from *Kr* to *pdm*, **(C)** activation from *pdm* to *cas*, **(D)** repression from *hb* to *cas*, and **(E)** the minimum network. The parameters involved in minimum network are shown. Each spoke represents a value of indicated parameter between the range used for random parameter assignment (Table IV). The value of $\tau_{x,\,off}$ is shown by normal scale and those of the other parameters are shown by log scale. Each polygon indicates one parameter set. Solid and broken lines indicate mean and s.d. of obtained parameters. The data are drawn from 5,000 trials of the random assignment of parameter values.

**Figure 8. Parameter dependencies of robustness for the *Drosophila* network.**

The fractions of successes for random assignment of parameter values are plotted under the different strengths of regulations ($\tilde{J}_{pdm, Kr}$, $\tilde{J}_{cas, pdm}$, and $\tilde{J}_{cas, hb}$) and default promoter activities ($S_{pdm}$ and $S_{cas}$). Dependencies of robustness to **(A)** $\tilde{J}_{pdm, Kr}$ (strength of activation from *Kr* to *pdm*) and $S_{pdm}$, **(B)** $\tilde{J}_{cas, pdm}$ (strength of activation from *pdm* to *cas*) and $S_{cas}$, and **(C)** $\tilde{J}_{cas, hb}$ (strength of the repression from *hb* to *cas*) and $S_{cas}$. The other parameters are set as in Table IV. The temporal dynamics were checked for 50,000 trials.

**Figure 9. Regulatory module for the precise sequential expression.**

The regulatory interactions and schematic expression profiles of the networks. **(A)** Regulatory interactions of the minimum network for sequential expression (left). This network reproduces the sequential expression under appropriate conditions (middle). However the parameter variations from the appropriate values and the increase of noise could easily alter the expression profiles (right). **(B)** Regulatory interactions of the *Drosophila* network (left). Three different kinds of regulations in this network enable the temporal expression in the precise order.

**Tables**

**Table I. List of the regulatory interactions of the genes in the NB temporal patterning network**

| Regulations | | References |
|---|---|---|
| Activation | *hb* → *Kr* | [22] |
| | *Kr* → *pdm* | [22] |
| | *pdm* → *cas* | [23] |
| Repression | *hb* ⊣ *pdm* | [21], [22] |
| | *hb* ⊣ *cas* | [23] |
| | *Kr* ⊣ *cas* | [22] |
| | *pdm* ⊣ *Kr* | [23] |
| | *cas* ⊣ *pdm* | [21], [23] |
| | *svp* ⊣ *hb* | [25], [28] |

**Table II. List of references for the sequential expression pattern in the wild type and mutants of the genes in the NB temporal patterning network**

| Genotype | References |
|---|---|
| wt | [22], [25] |
| *hb*⁻ | [22], [27] |
| *Kr*⁻ | [22] |
| *pdm*⁻ | [23], [32] |
| *cas*⁻ | [23], [32] |
| *hb* o.e.[1] | [22] |
| *Kr* o.e. | [22], [31] |
| *pdm* o.e. | [23], [32] |
| *cas* o.e. | [23], [32] |

1. o.e.: over expression.

**Table III. Criterion for expression profile in each genotype**

| Genotype | Criterion for the expression profile |
|---|---|
| wt | $\tau_{hb,on(off)} \leq \tau_{kr,on(off)} \leq \tau_{pdm,on(off)} \leq \tau_{cas,on(off)}$, $(\tau_{i,on} \neq \tau_{j,on}) \cup (\tau_{i,off} \neq \tau_{j,off})$ |
| $hb^-$ | $\tau_{kr,on(off)} \leq \tau_{pdm,on(off)} \leq \tau_{cas,on(off)}$, $(\tau_{i,on} \neq \tau_{j,on}) \cup (\tau_{i,off} \neq \tau_{j,off})$ |
| $Kr^-$ | $\tau_{hb,on(off)} \leq \tau_{pdm,on(off)} \leq \tau_{cas,on(off)}$, $(\tau_{i,on} \neq \tau_{j,on}) \cup (\tau_{i,off} \neq \tau_{j,off})$ |
| $pdm^-$ | $\tau_{hb,on(off)} \leq \tau_{kr,on(off)} \leq \tau_{cas,on(off)}$, $(\tau_{i,on} \neq \tau_{j,on}) \cup (\tau_{i,off} \neq \tau_{j,off})$ |
| $cas^-$ | $\tau_{hb,on(off)} \leq \tau_{kr,on(off)} \leq \tau_{pdm,on(off)}$, $(\tau_{i,on} \neq \tau_{j,on}) \cup (\tau_{i,off} \neq \tau_{j,off})$ |
| hb o.e. | $\tau_{hb,on(off)} \leq \tau_{kr,on(off)}$, $(\tau_{i,on} \neq \tau_{j,on}) \cup (\tau_{i,off} \neq \tau_{j,off})$, $X^t_{pdm} = X^t_{cas} = 0$ |
| Kr o.e. | $\tau_{hb,on(off)} \leq \tau_{kr,on(off)} \leq \tau_{pdm,on(off)}$, $(\tau_{i,on} \neq \tau_{j,on}) \cup (\tau_{i,off} \neq \tau_{j,off})$, $X^t_{cas} = 0$ |
| pdm o.e. | $\tau_{hb,on(off)} \leq \tau_{pdm,on(off)} \leq \tau_{cas,on(off)}$, $(\tau_{i,on} \neq \tau_{j,on}) \cup (\tau_{i,off} \neq \tau_{j,off})$, $X^t_{kr} = 0$ |
| cas o.e. | $\tau_{hb,on(off)} \leq \tau_{kr,on(off)}$, $(\tau_{i,on} \neq \tau_{j,on}) \cup (\tau_{i,off} \neq \tau_{j,off})$, $X^t_{pdm} = 0$ |

**Table IV. Parameter values used for continuous dynamics of the genetic networks**

| Parameter | Biological meaning | Value |
|---|---|---|
| $\gamma_M$ | Degradation rate of mRNAs | 1.0 |
| $\gamma_P$ | Degradation rate of proteins | 0.2 |
| $\tau_{hb,off}$ | Time for promoter activity of *hb* switched off | 10.0 |
| $\tau_{x,off}$ | Time for promoter activity of *x* switched off | $[0.5\tau_{hb,off}, 2.0\tau_{hb,off}]$ |
| $K_M$ | Michaelis constant for the promoter functions | 0.1 |
| $\alpha$ | Hill coefficient for the promoter functions | 2.0 |
| $\tilde{J}_{ij}$ | Strength of regulation from gene *j* to gene *i* | $|\tilde{J}_{ij}| \in [10^{-1}, 10^{1}]$ |
| $S_i$ | Default promoter activity of gene *i* | $S_{i\,(\neq hb)} \in [10^{-3}, 10^{1}]$, $S_{hb} \in [2 \times 10^{-1}, 10^{1}]$ |

# Figures

## Figure 1.

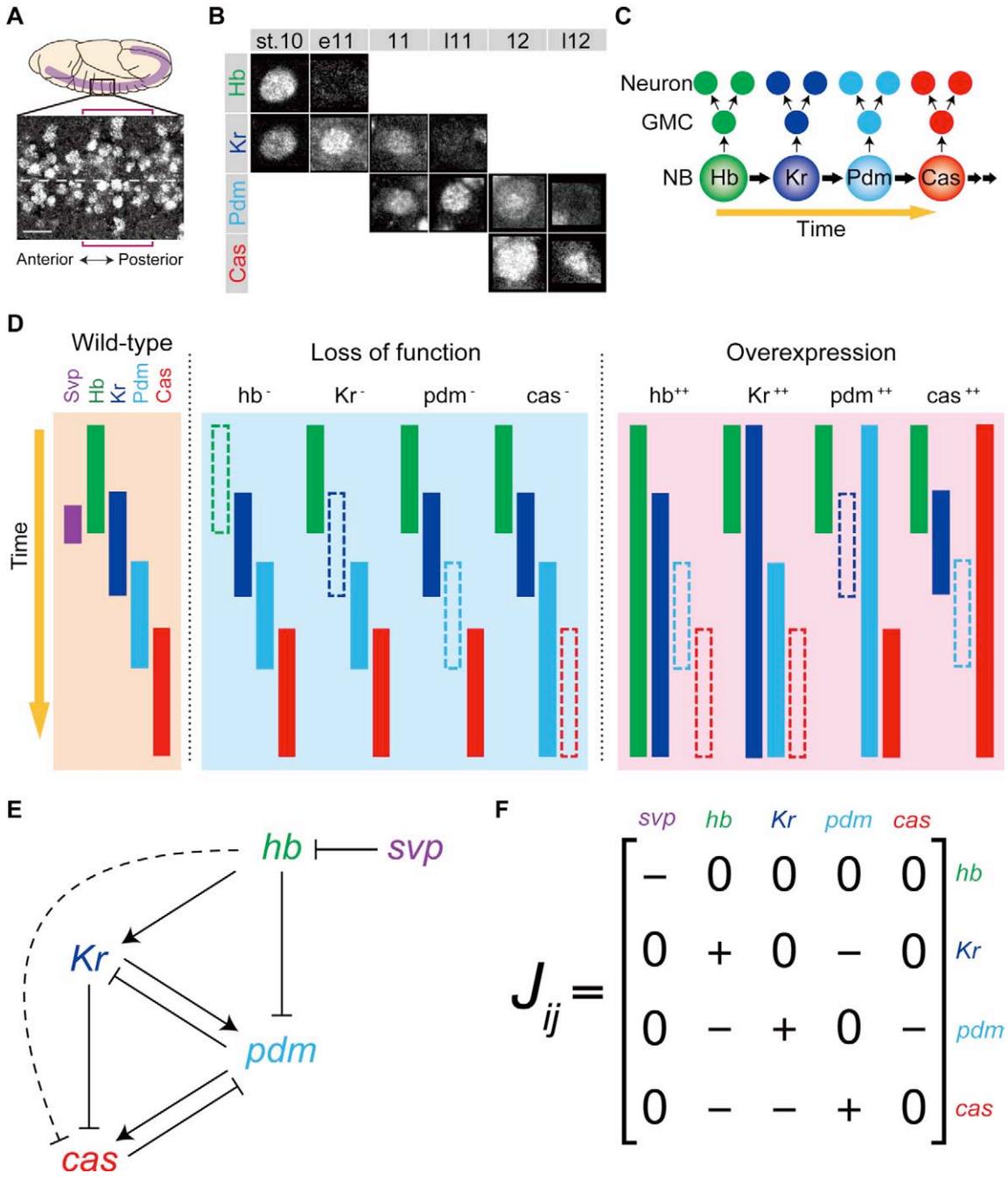

**Figure 2.**

**A** $\theta_{Kr}=0$, $\theta_{pdm}=0$, $\theta_{cas}=1$

**B** $\theta_{Kr}=0$, $\theta_{pdm}=1$, $\theta_{cas}=1$

**C** $\theta_{Kr}=1$, $\theta_{pdm}=1$, $\theta_{cas}=1$

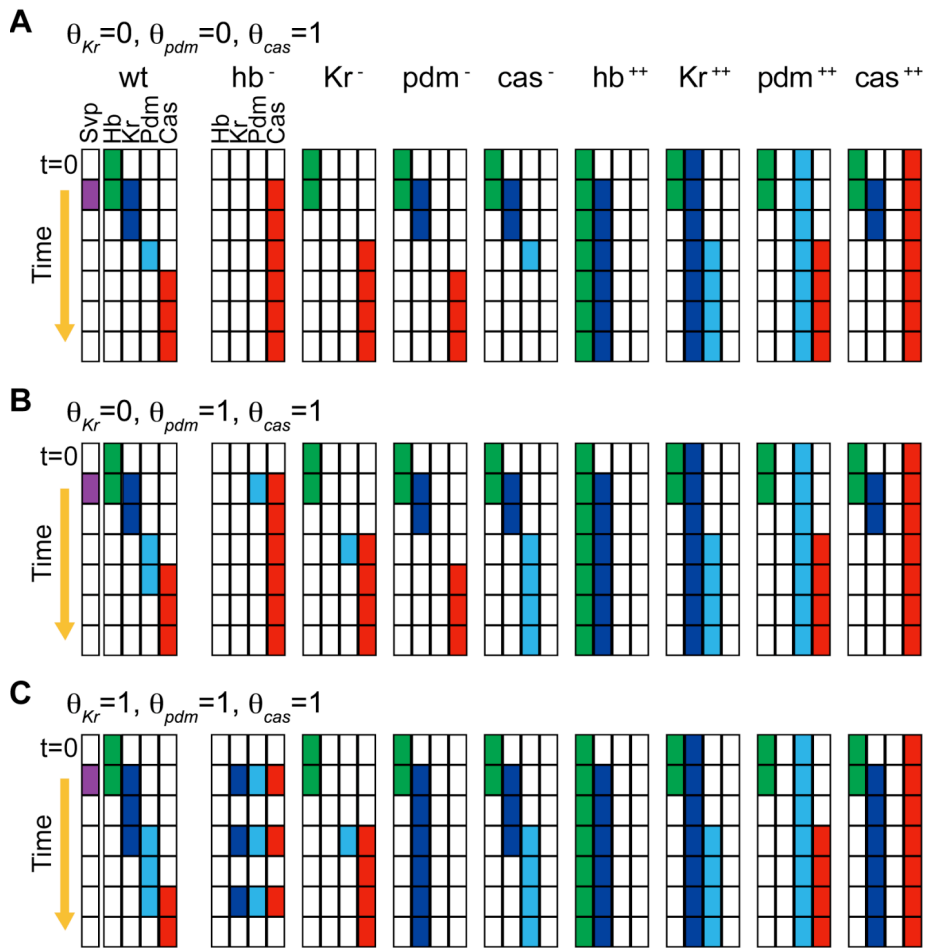

**Figure 3.**

A

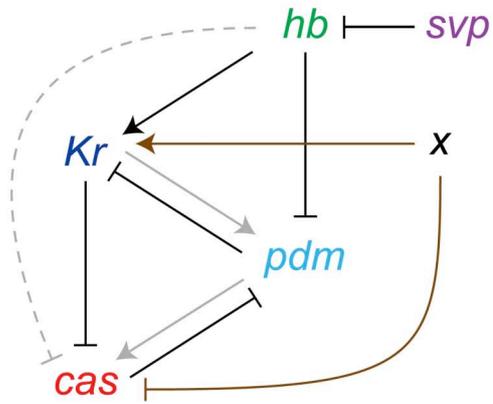

B

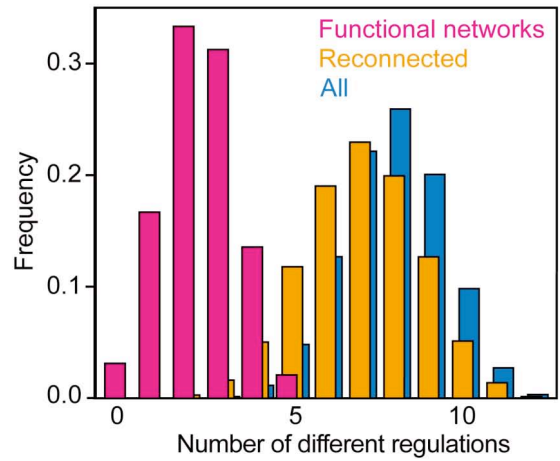

C

**Figure 4.**

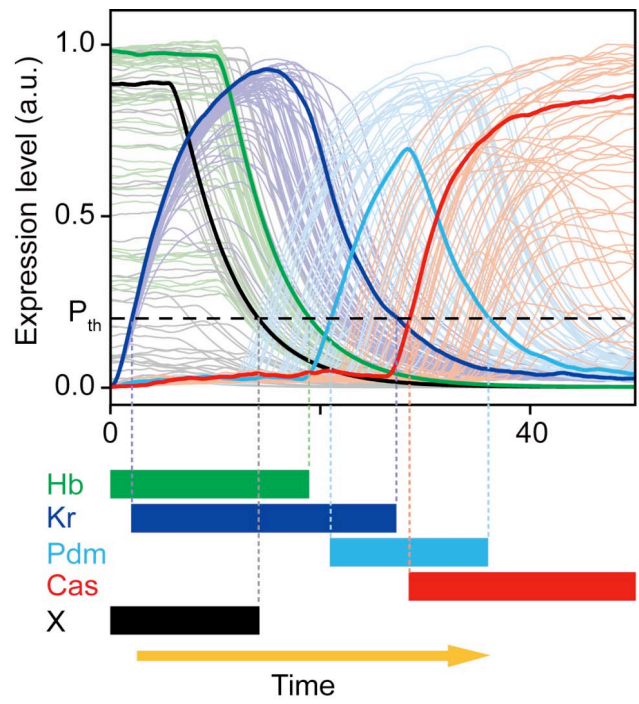

**Figure 5.**

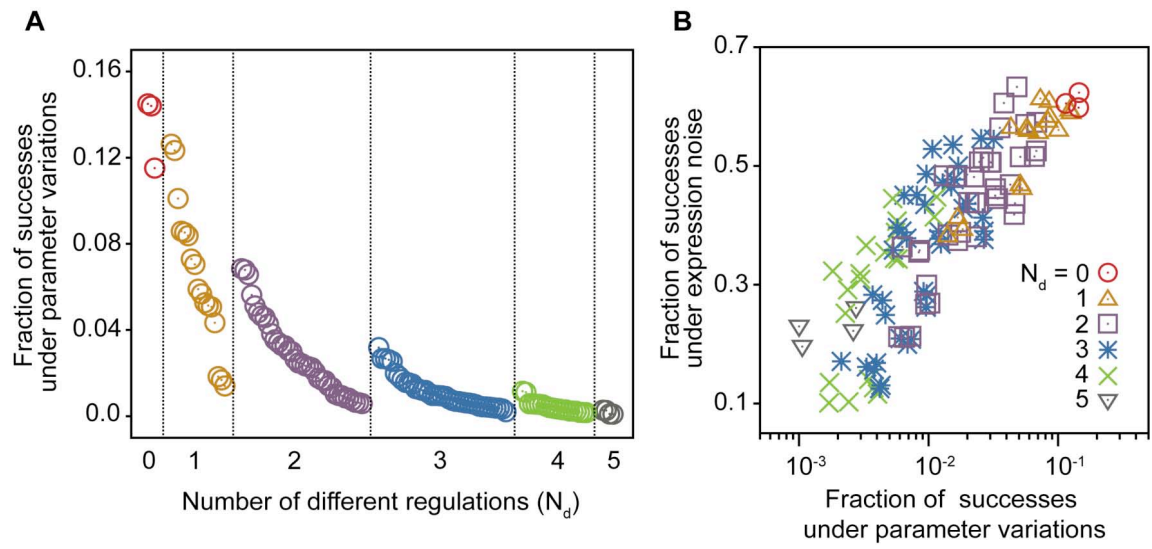

**Figure 6.**

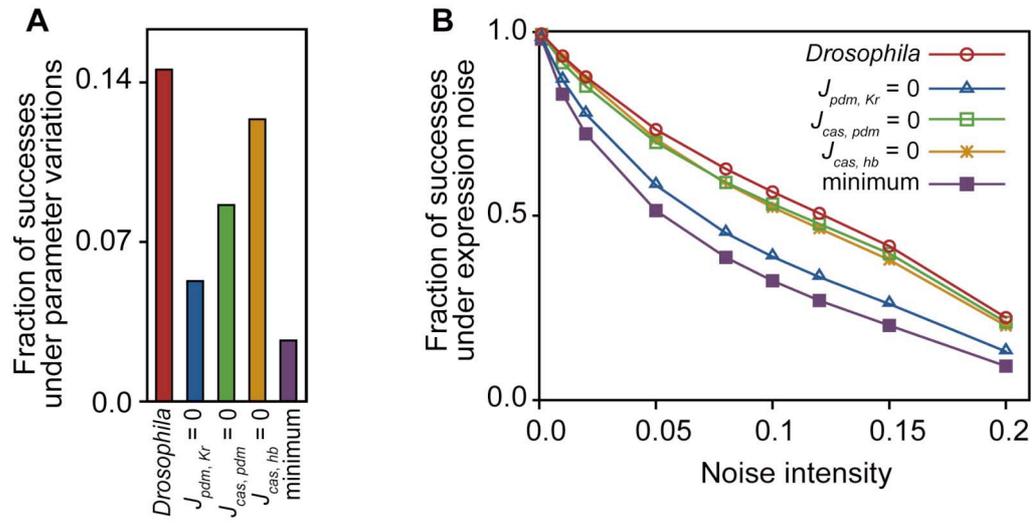

**Figure 7.**

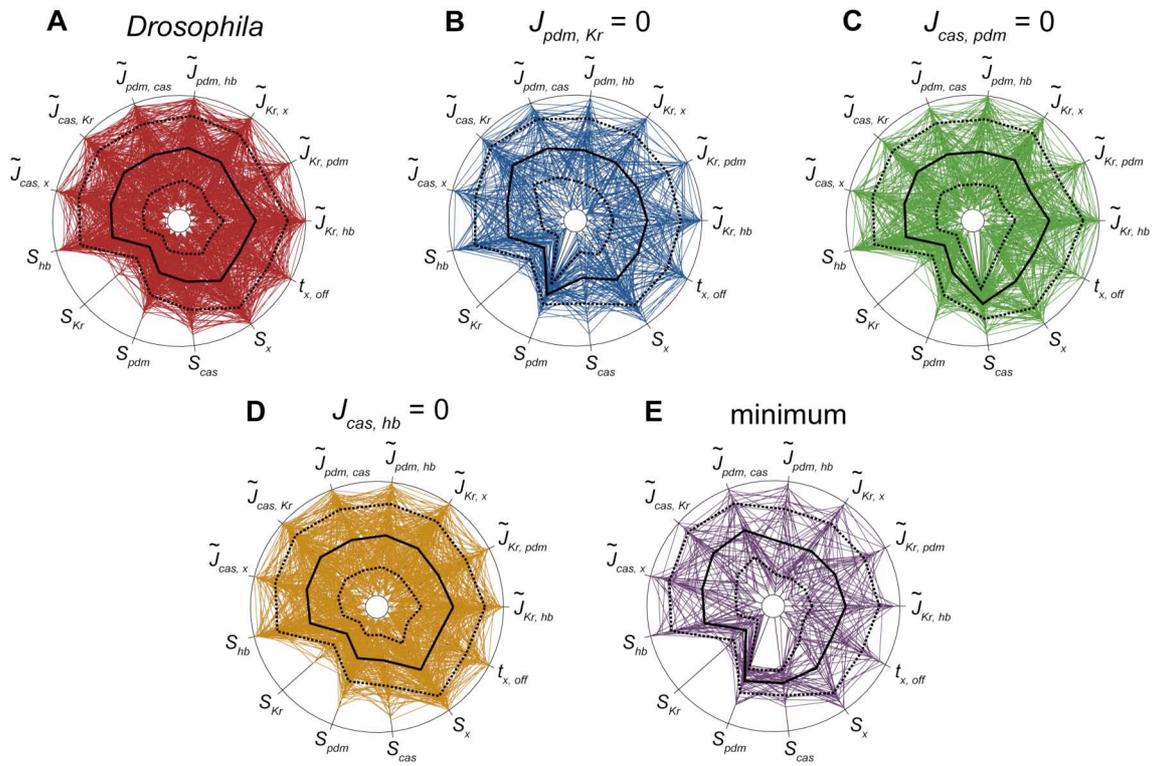

**Figure 8.**

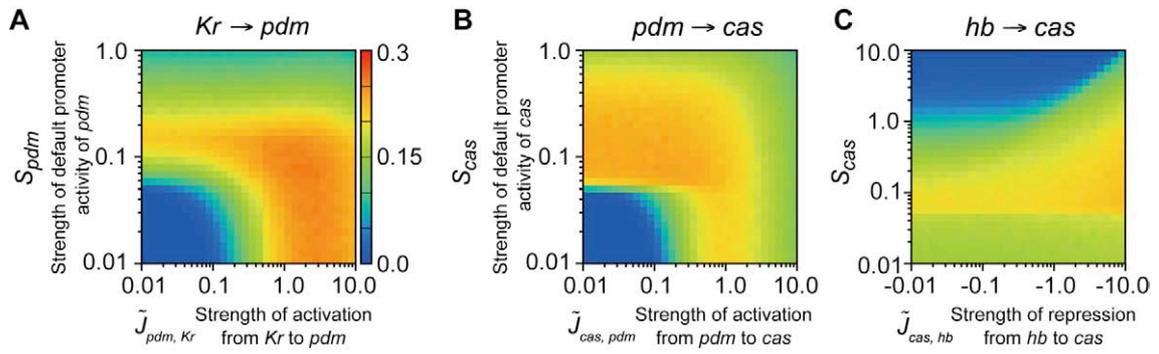

**Figure 9.**

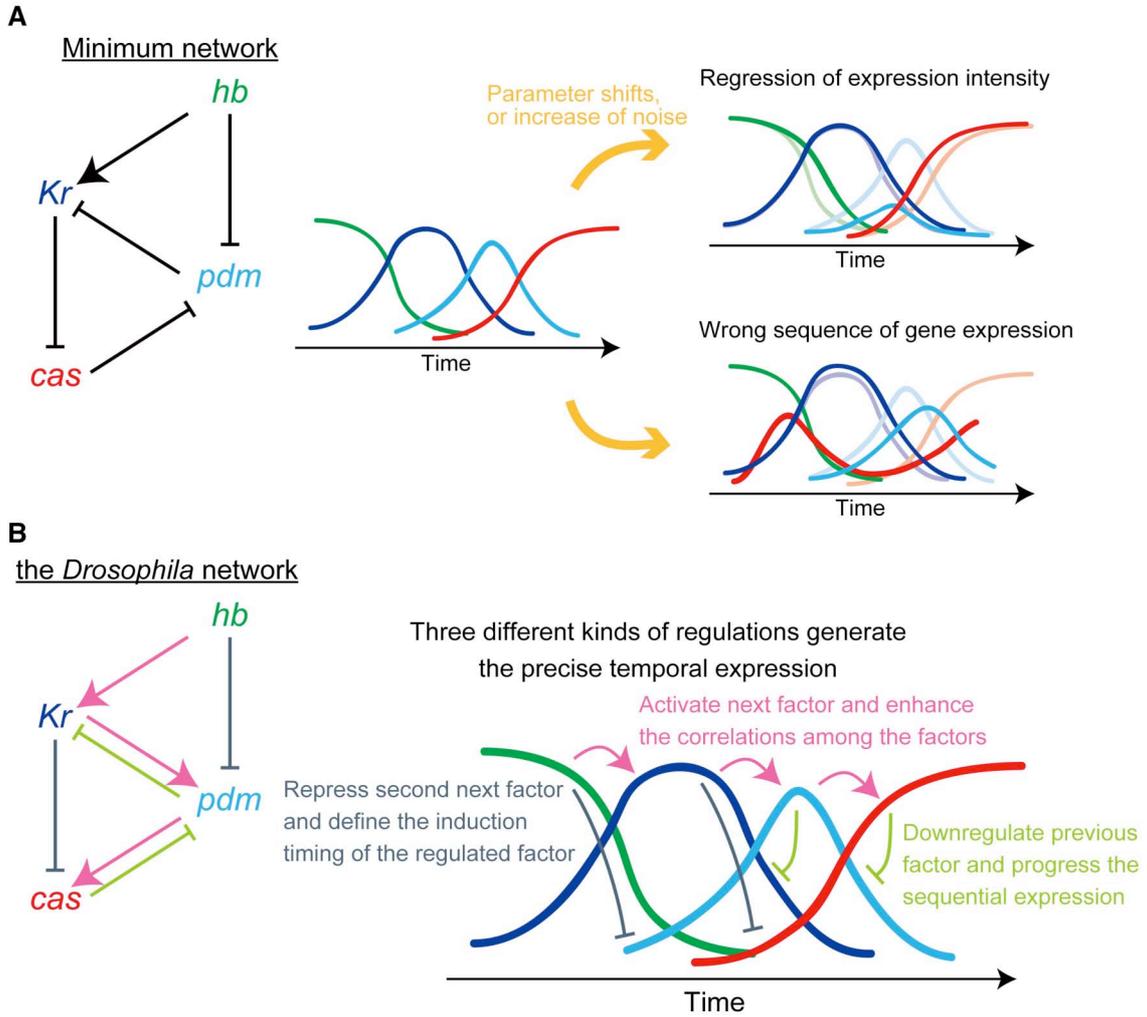